\documentclass{PoS}

\title{Efficient approximations of neutrino physics for three-dimensional simulations of stellar core collapse}

\ShortTitle{Efficient approximations for 3D simulations of stellar core collapse}

\author{\speaker{Matthias Liebend\"orfer}\\
        University of Basel\\
        E-mail: \email{matthias.liebendoerfer@unibas.ch}}

\author{Ue-Li Pen, Christopher Thompson\\
        CITA, University of Toronto\\
        E-mail: \email{pen@cita.utoronto.ca, thompson@cita.utoronto.ca}}

\abstract{
Neutrino transport in spherically symmetric models of stellar core
collapse and bounce has achieved a technically complete level,
rewarded by the agreement among independent groups that a
multi-dimensional treatment of the fluid-instabilities in the
post-bounce phase is indispensable to model supernova explosions.
While much effort is required to develop a reliable neutrino
transport technique in axisymmetry, we explore neutrino physics
approximations and parameterizations for an efficient three-dimensional
simulation of the fluid-instabilities in the shock-heated matter
that accumulates between the accretion shock and the protoneutron
star. We demonstrate the reliability of a simple parameterization
scheme in the collapse phase and extend our 3D magneto-hydrodynamical
collapse simulations to a preliminary postbounce evolution.
The growth of magnetic fields is investigated.
}

\FullConference{International Symposium on Nuclear Astrophysics - Nuclei in the Cosmos - IX\\
		 25-30 June 2006\\
		 CERN}

\begin{document}

\begin{figure}[t]
  \begin{minipage}{0.45\textwidth}
    \includegraphics[width=\textwidth]{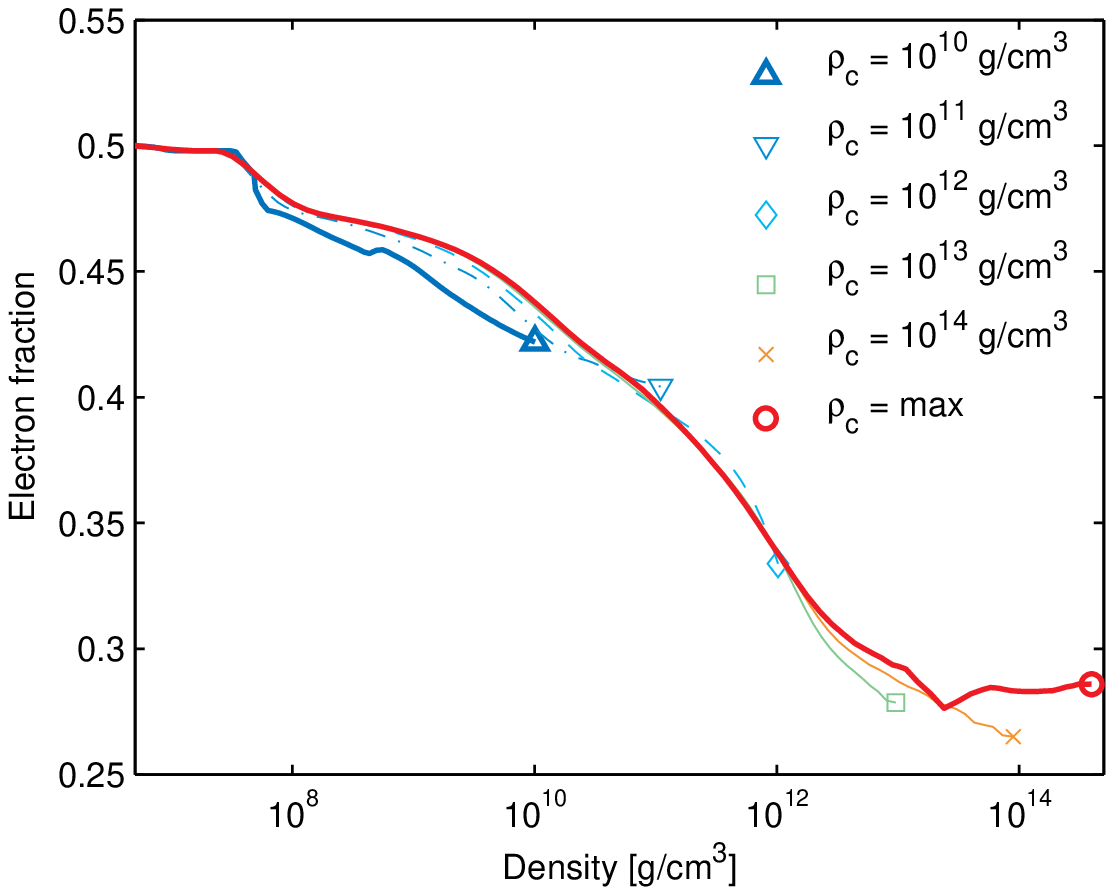}
    \caption{Electron fraction profiles during core collapse in model G15
      \protect\cite{Liebendoerfer.Rampp.ea:2005}. Each line shows \( Y_e \)
      as a function of density at a given time. As \( Y_e(\rho ) \) is only
      a weak function of time and because the accuracy is most important
      at core-bounce, the parameterization of \( Y_e \) is
      conveniently based on the bounce profile labelled by a circle.}
    \label{fig1.eps}
  \end{minipage}
  \hspace{\fill}
  \begin{minipage}{0.45\textwidth}
    \center{\includegraphics[width=0.7\textwidth]{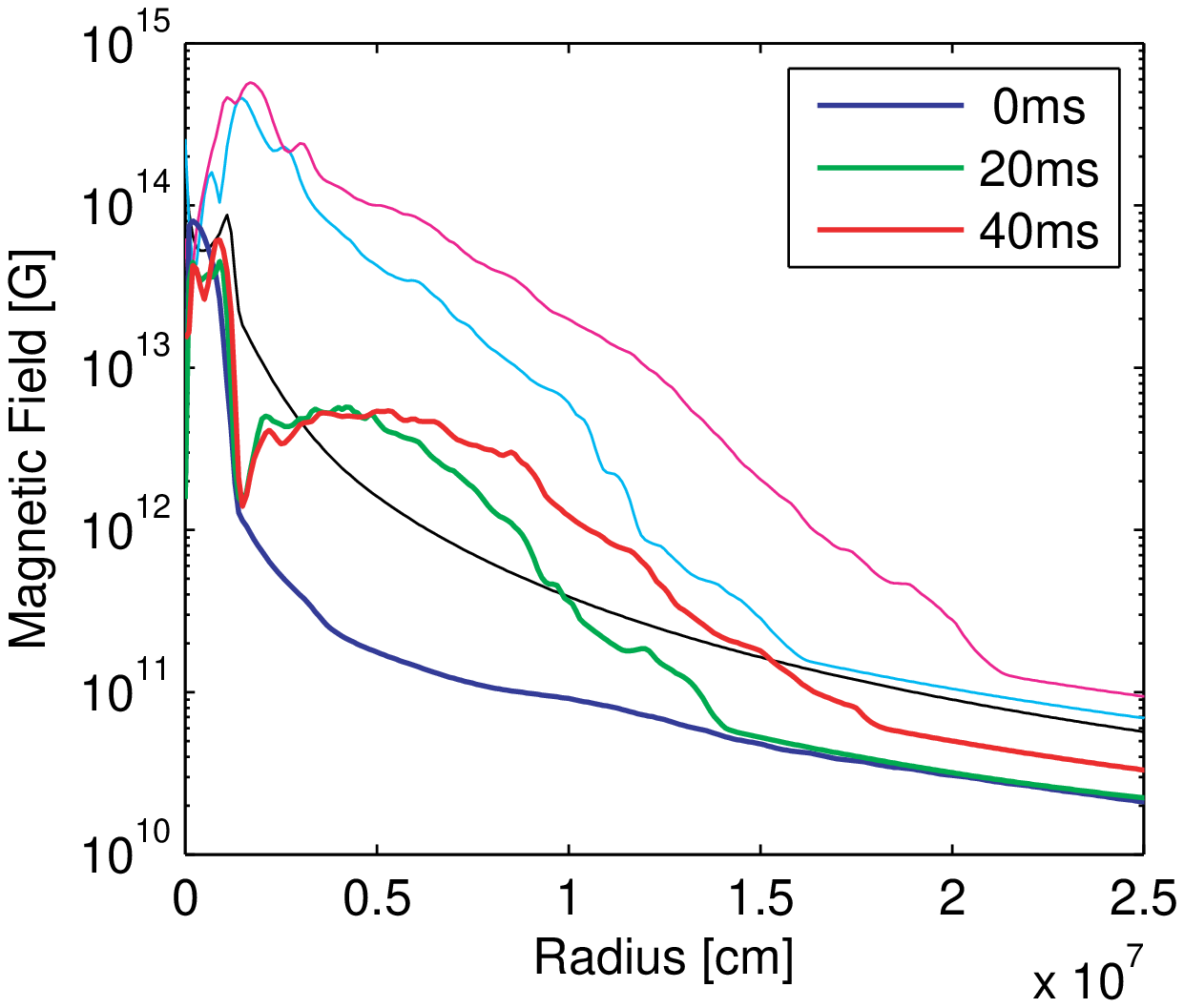}}
    \caption{Strength of the magnetic field at \( 0 \) ms, \( 20 \) ms, and
      \( 40 \) ms after bounce in two different models.
      Model 1 (black, cyan, magenta), which starts with
      a poloidal field and \( \Omega=2\pi \) rad/s, shows rapid field
      amplification by winding at the surface of the protoneutron star.
      This effect is negligible in model 2 (blue, green, red), which started
      with a toroidal field and \( \Omega=0.05 \) rad/s.}
    \label{fig2.eps}
  \end{minipage}
\end{figure}
\section{Parameterization of neutrino physics and comparison with reference data}
In multi-dimensional simulations of stellar core collapse,
rotating progenitors, general relativity, and magnetic fields
have been investigated.
Only few recent multi-dimensional collapse simulations made the effort
to include neutrino physics. These schemes are either very
computationally expensive \cite{Buras.Rampp.ea:2003,Dessart.Burrows.ea:2006}
or rely on simplifications of the neutrino transport and its microphysics
\cite{Kotake.Yamada.Sato:2003,Fryer.Warren:2004}.
The dynamics of core collapse is dominated by electron pressure.
Dynamical simulations are only realistic if they take
electron captures into account. At
increasing densities, the electron captures get inhibited by neutrino
phase space blocking so that the ability to thermalize and emit
the produced neutrinos significantly contributes to the determination
of the final electron fraction (\( Y_e \)) in the inner core.  The lower
the \( Y_e \), the smaller is the mass of the core that bounces
when nuclear densities are reached and the smaller is also the
initial energy imparted to the outgoing shock. These well-known
relationships ask for a careful inclusion of neutrino physics in
simulations of stellar core collapse. The direct solution of the
Boltzmann neutrino transport equation, however, has only been achieved
under the assumption of spherical symmetry (Ref. \cite{Mezzacappa:2005}
and references therein).

\begin{figure}[t]
  \center{\includegraphics[width=0.7\textwidth]{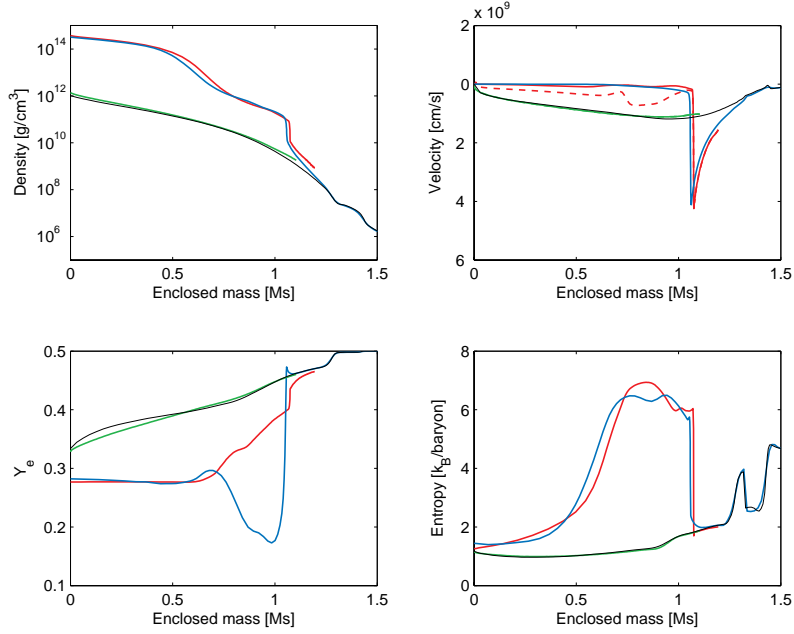}}
  \caption{Comparison of the outcome of a parameterized 3D
    simulation with model G15, which is based on general relativistic three-flavour
    Boltzmann neutrino transport \protect\cite{Liebendoerfer.Rampp.ea:2005}.
    From the upper left to the lower right we compare as a function of enclosed
    mass: the density-, the velocity-, the \( Y_e \)-, and the entropy profiles.
    The green and red lines represent the spherically averaged profiles from
    the 3D run at \( 5 \) ms before and \( 5 \) ms after bounce, respectively.
    The black and blue lines represent the G15 reference data at corresponding
    times. Excellent agreement is found in all four quantities---with one
    exception: This simple parameterized neutrino leakage cannot model the
    neutrino burst. The neutrino burst causes a prominent
    \( Y_e \)-dip and additional cooling in the G15 data.}
  \label{fig3.eps}
\end{figure}

Three-dimensional models must rely on approximations of the
neutrino transport to be treatable by current computers.
A simple and computationally efficient parameterization of the
deleptonization in the collapse phase
is based on a tabulation of \( Y_e \) as a function of
density. A spherically symmetric model with Boltzmann neutrino
transport \cite{Liebendoerfer.Rampp.ea:2005} shows that this function
is fairly independent of time so that it can be used throughout
the collapse phase (see Fig.~\ref{fig1.eps}). Once the deleptonization is
known, an estimate of entropy changes and neutrino stress can be deduced
\cite{Liebendoerfer:2005}.
The comparison of the full-domain 3D parameterized
runs with the spherically symmetric general relativistic reference
model G15 shows
that the 1D results of core collapse are accurately reproduced by
a 3D run if it is launched from a non-rotational progenitor (see Fig.~\ref{fig3.eps}). The parameterized
neutrino physics presents a significant improvement with respect
to adiabatic simulations and may even rival with
neutrino transport schemes that neglect neutrino-electron scattering.
However, the accuracy breaks down with the launch of the neutrino
burst at a few milliseconds after bounce. With the currently implemented
scheme, accretion flows in the postbounce phase deleptonize only down to
\( Y_e\sim 0.3 \) instead of \( Y_e\sim 0.15 \). Neutrino heating is
neglected altogether. Better neutrino physics approximations are required after
core collapse.

\section{Three-dimensional simulations with magnetic fields}
\begin{figure}[t]
  \center{\includegraphics[width=0.85\textwidth]{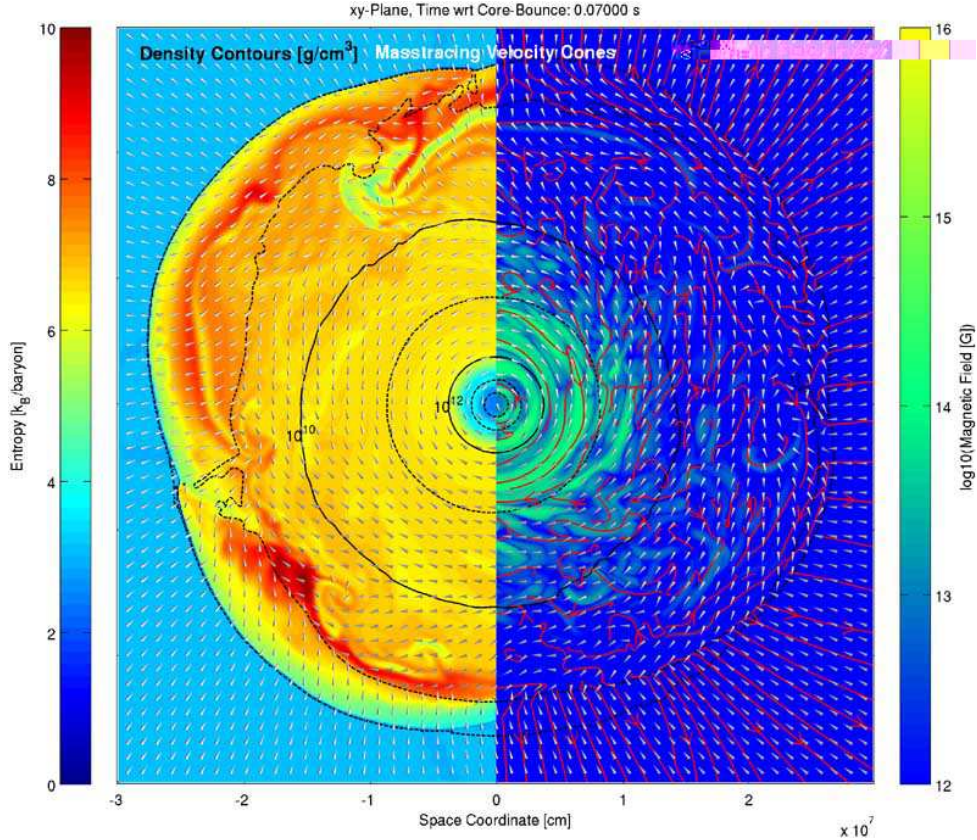}}
  \caption{Slice through the equatorial plane of a 3D simulation at
    \( 70 \) ms after bounce. On the left hand side the background color
    indicates the entropy per baryon, and on the right hand side the magnetic
    field strength. Density contours are drawn in black, and magnetic field
    lines in red. The white cones align with the 3D velocity field. In this
    rotating model, instabilities of the standing accretion front develop
    more strongly in the equatorial plane than in the polar direction.}
  \label{fig4.eps}
\end{figure}
The 3D simulations
are based on a simple and fast  cosmological MHD
code \cite{Pen.Arras.Wong:2003} which has been parallelized, improved
and adapted to the supernova context. A realistic equation of state
\cite{Lattimer.Swesty:1991} is used and gravity is implemented
by a spherically symmetric mass integration which includes general
relativistic corrections \cite{Marek.Dimmelmeier.ea:2006}.
The simulations are launched from a \( 15 \) M$_{\odot}$ progenitor
model \cite{Woosley.Weaver:1995}.
The 3D computational domain consists of a central region of \( 600 \) km\(^3\),
treated in equidistant Cartesian coordinates with a resolution of
\( 1 \) km. Outside of the neutron star, this resolution is comparable to
or better than the resolution chosen in 1D simulations (cf.
the shock widths in Fig.~\ref{fig3.eps}). The 3D computational domain
is embedded in a larger spherically symmetrical computational domain treated
by a 1D hydrodynamics code. In the pre-collapse configuration, an initial
angular momentum is assigned to the spherical mass shells according to
an angular velocity of \( 2\pi \) rad/s and a shellular quadratic cutoff at
\( 500 \) km radius. The angular momentum is assumed to be conserved
until the infalling layers enter the 3D computational domain.
The initial poloidal magnetic field is derived from a vector
potential whose norm scales with the square root of the density. It
is set to produce a field strength of \( 5\times 10^9 \) Gauss
at a reference density of \( 5\times 10^7 \) g/cm$^3$.

As the duration of the collapse phase is no longer than a fraction
of a rotational period, the magnetic field does not significantly
wind up during collapse. Its amplification by about two orders of
magnitude is due to the compression of field lines in the condensing
matter. A poloidal initial field develops a biconical
shape, focussed on the protoneutron star at the origin. Toroidal
initial fields tend to remain toroidal throughout collapse.
Figure \ref{fig2.eps} compares a
fast rotating model with poloidal field to a model where initial
rotation and magnetic fields are set to the values suggested in
Ref. \cite{Heger.Woosley.Spruit:2005}. A different amplification
of the field strength by winding becomes evident close to the
surface of the protoneutron star. The offset between the field
strengths of simultaneous profiles at larger distances, however,
stem from the initial configuration\footnote{The vector potentials
used to generate the initial fields align differently with the
gradient of the density to which they were scaled. Thus,
the toroidal case starts with a smaller field strength
in the outer layers than the poloidal case.}.
After bounce, the declining strength of the bounce-shock results in a
negative entropy gradient. The hot matter layered around the
protoneutron star becomes convectively unstable on a short time
scale. This causes the magnetic field lines in this region to
entangle (see Fig.~\ref{fig4.eps}) and to grow further.

\section*{Acknowledgments}
This work is funded by the Swiss National Science Foundation,
grant No. PP002-106627. The simulations have been carried out on
the Athena cluster at the University of Basel and on the McKenzie cluster
\cite{Dubinski.Humble.ea:2003} which was funded by the Canada Foundation
for Innovation and the Ontario Innovation Trust.

\providecommand{\href}[2]{#2}\begingroup\raggedright
\endgroup

\end{document}